\documentclass[11pt,a4paper]{article}
\pdfoutput=1
\usepackage{jcappub}
\newcommand{\be}{\begin{equation}}
\newcommand{\ee}{\end{equation}}
\newcommand{\Li}{\mathop{\mathrm{Li}}\nolimits}

\title{Universal Doomsday: Analyzing Our Prospects for Survival}

\author[a]{Austin Gerig,}

\author[b]{Ken D. Olum,}

\author[b]{and Alexander Vilenkin}

\affiliation[a]{CABDyN Complexity Centre, Sa\"{i}d Business School, University of
Oxford, Oxford OX1 1HP, UK}
\affiliation[b]{Institute of Cosmology, Department of Physics and Astronomy,
Tufts University, Medford MA 02155, USA}

\emailAdd{austin.gerig@sbs.ox.ac.uk}
\emailAdd{kdo@cosmos.phy.tufts.edu}
\emailAdd{vilenkin@cosmos.phy.tufts.edu}

\abstract{Given a sufficiently large universe, numerous civilizations
  almost surely exist.  Some of these civilizations will be
  \emph{short-lived} and die out relatively early in their
  development, i.e., before having the chance to spread to other
  planets.  Others will be \emph{long-lived}, potentially colonizing
  their galaxy and becoming enormous in size.  What fraction of
  civilizations in the universe are long-lived?  The ``universal
  doomsday'' argument states that long-lived civilizations must be
  rare because if they were not, we should find ourselves living in
  one.  Furthermore, because long-lived civilizations are rare, our
  civilization's prospects for long-term survival are poor.  Here, we
  develop the formalism required for universal doomsday calculations
  and show that while the argument has some force, our future is not
  as gloomy as the traditional doomsday argument would suggest, at
  least when the number of early existential threats is small.}


\begin{document}

\maketitle

\section{Introduction}

The Doomsday Argument
\cite{Carter:unpublished,Leslie:1989}\footnote{Gott
  \cite{Gott:doomsday} and Nielsen \cite{Nielsen:doomsday} make
  similar arguments.} traditionally runs as follows.  Perhaps our
civilization will soon succumb to some existential threat (nuclear
war, asteroid impact, pandemic, etc.)~so that the number of humans
ever to exist is not much more than the number who have existed so
far.  We will call such a civilization \emph{short-lived} and the
total number of humans in it $N_S$.  Alternatively, we might survive
all such threats and become \emph{long-lived}, potentially colonizing
other planets and eventually generating a large total number of
individuals, $N_L$.  For simplicity we will consider only two possible
sizes.  We expect $N_L\gg N_S$.  The ratio $R= N_L/N_S$ could easily
be as large as a billion.

We don't know our chances of being short- or long-lived, but we can assign some prior belief or credence $P(S)$ and $P(L) = 1-P(S)$ in these
two possibilities.  These confidence levels should be based on our
analysis of specific threats that we have considered and the
possibility of other threats of which we are not yet aware.

Suppose that you hold such confidence levels at a time when you don't
know your own position in the entire human race.  Now you discover
that you are one of the first $N_S$ humans to be born.  We will call
this datum $D$.  If the human race is to be short-lived, $D$ is
certain.  If the human race is to be long-lived, assuming that you can
consider yourself a randomly chosen human
\cite{Vilenkin:1995ua,Page:1995kw,Bostrom:book,Garriga:2007wz}, the chance that
you would be in the first $N_S$ is only $1/R$.  Thus you should update
your probabilities using Bayes' Rule, to get
\be\label{eq.traditional}
P(L|D) = \frac{P(L)/R}{P(L)/R + P(S)} = \frac{P(L)}{P(L) + P(S)R} < \frac{1}{P(S)R}\,.
\ee
Since it is clear that we do face existential threats, $P(S)$ is not
infinitesimal.  Thus $P(S) R\gg 1$, $P(L|D)\ll1$, and doom (our
civilization ending soon rather than growing to large size) is nearly
certain.

Many counterarguments have been offered (e.g.,
\cite{Dieks:sia,Page:sia,Bartha:sia,Bostrom:book,Olum:2000zw}). The
specific issue which will concern us here is the possibility that our
universe might contain many civilizations.  In that case, we should
consider ourselves to be randomly chosen from all individuals in that
universe or multiverse.  Before taking into account $D$, our chance to
be in any given long-lived civilization is then higher than our chance
to be in any given short-lived civilization by factor $R$.  Taking
into account $D$ simply cancels this factor, so the chance that we are
in a long-lived civilization is just the fraction of civilizations that
are long-lived.  (More compactly, since each civilization contains $N_S$
individuals who observe $D$, this observation provides no reason to
prefer one type of civilization over another.)

Thus if there are many civilizations, the doomsday argument is
defeated.  However, it returns in another form, called the
\emph{universal doomsday argument} \cite{Knobe:2003js,Gerig:2012qg}.
We are more likely to observe $D$ in a universe in which most
civilizations are short-lived.  While we can no longer conclude
anything unique to our own civilization, we can conclude that most
civilizations are likely to be short-lived, and thus ours is likely to
be short-lived also.

This paper analyzes the universal doomsday argument.  In contrast to
the traditional doomsday argument, which in almost all circumstances
makes doom nearly certain, for many reasonable priors the universal
argument gives only a mildly pessimistic conclusion.  However, for
other priors, the conclusion of the universal argument can be quite
strong.

The analysis of many civilizations in the universe can be extended to
analyze possible civilizations that might exist according to different
theories of the universe.  If we take ourselves to be randomly chosen
among all observers that might exist \cite{Dieks:sia,Olum:2000zw}, the
doomsday argument is completely canceled, and after taking into
account $D$ we find no change in our prior probabilities for different
lifetimes of our civilization.  This assumption, equivalent to the
\emph{self-indication assumption} \cite{Dieks:sia}, is a controversial
one, and the authors of the present paper are not in agreement about
it.  However, for the purpose of the present work we will consider the
consequences of denying this idea, and consider ourselves to be
randomly chosen only among those individuals who exist in our actual
universe.

In the next section we set up the formalism for calculating the
probability $P(L)$ for a civilization to be long-lived.  To simplify
the discussion, we focus on a special case where civilizations can
have only two possible sizes; the general case is discussed in the
Appendix.  In Sections~\ref{sec:firstprior}--\ref{sec:lastprior} this
formalism is applied to find $P(L)$ for several choices of the prior.
Our conclusions are summarized and discussed in Section \ref{sec:summary}.

\section{The fraction of long-lived civilizations in the universe}
\label{sec:setup}

Assume the universe is large enough such that numerous civilizations
exist.  Assume furthermore that a fraction $f_L$ of the civilizations
are long-lived, containing $N_L$ individuals each, and the remaining
fraction $1-f_L$ of the civilizations are short-lived, containing only
$N_S$ individuals each.  (The general case in which civilizations may
have any size, rather than just the specific sizes $N_S$ and $N_L$, is
discussed in the appendix.)  We will take the universe to be finite or
the problems that arise in infinite universes to have been solved, so
that the fraction $f_L$ is well defined.

We do not know what $f_L$ is, but we know it must be in the interval,
$[0,1]$.  We can represent our prior belief, or ``best guess'' for
different values of $f_L$, with a density function $P(f_L)$, so that
$P(f_L) df_L$ is our prior probability for the
fraction of long-lived civilizations to lie within an infinitesimal
interval $df_L$ around $f_L$.  We only consider normalized priors so
that $\int_0^1 P(f_L)df_L =1$.

\begin{figure}
\begin{center}
\includegraphics[width=3.4in]{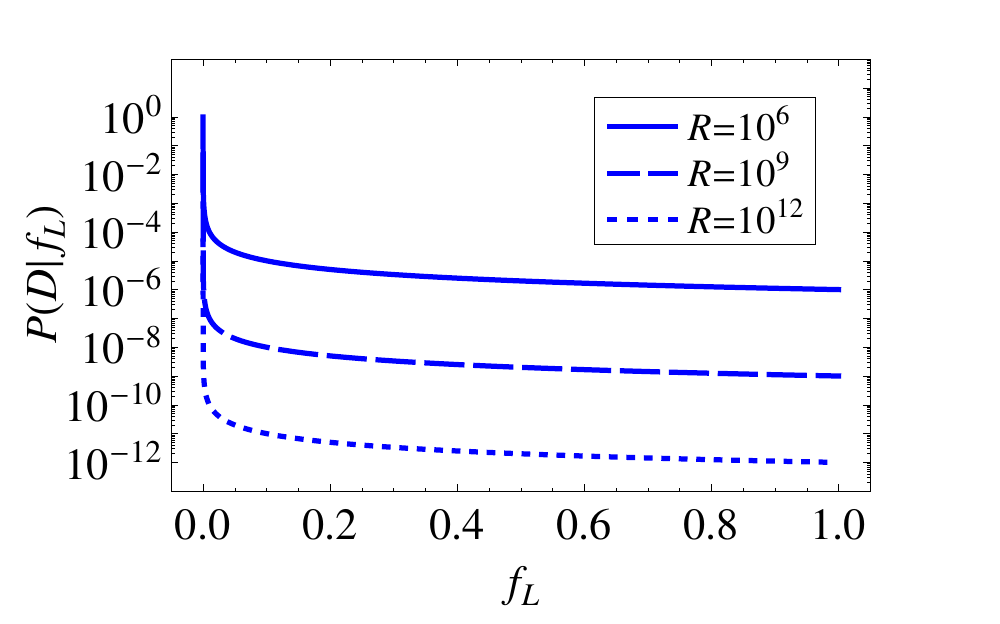}
\end{center}
\caption{The likelihood function, $P(D|f)$, given $R=10^6$, $R=10^9$, and $R=10^{12}$.}
\label{fig.likelihood}
\end{figure}

Whatever we take as our prior, we should update it after observing new
evidence or data.  The rest of this paper is concerned with updating
$P(f_L)$ after considering the datum $D$: we were born into a
civilization that has not yet reached long-lived status.  Bayes'
Rule gives
\begin{equation}
P(f_L|D) = \frac{P(D|f_L)P(f_L)}{\int df_L P(D|f_L)P(f_L)}\,.
\label{eq.bayes}
\end{equation}
With a given $f_L$, the probability of observing $D$, $P(D|f_L)$, is just the
ratio of the number of observers making observation $D$ to the total
number of observers.  Factoring out the total number of civilizations,
we divide the number of observers in each civilization finding $D$,
which is just $N_S$, by the average number of observers per
civilization, $N_Lf_L + N_S(1-f_L)$.  Thus
\be
P(D|f_L) = \frac{N_S }{N_S (1-f_L) + N_L f_L}
= \frac{1}{1+f_L(R-1)}\,. \label{eq.likelihood}
\ee
In Fig.~\ref{fig.likelihood}, we plot the likelihood function given $R=10^6$, $R=10^9$, and $R=10^{12}$.

Plugging the likelihood into \eqref{eq.bayes},
\begin{equation}
P(f_L|D) = \frac{P(f_L)/(1+f_L(R-1))}{\int_0^1 df_L P(f_L)/(1+f_L(R-1))}\,.
\label{eq.posterior}
\end{equation}
The posterior distribution, $P(f_L|D)$, expresses our updated belief
in different values of $f_L$ after considering that our civilization
is not yet long-lived.

Notice that $P(f_L|D) \propto P(D|f_L)P(f_L)$ and that $P(D|f_L)$ is
much larger for low values of $f_L$ than for high values of $f_L$ (see
Fig.~\ref{fig.likelihood}).  Therefore, when updating our prior, we
place more credence in low values of $f_L$ and less credence in high
values of $f_L$, and so become more pessimistic about the fraction of
civilizations that reach long-lived status.

We can now compute the probability that our civilization will
eventually be long-lived.  For any given $f_L$, this probability is
just $f_L$, the fraction of civilizations that are long-lived.
Without considering $D$, we would just take an average of the possible
$f_L$ weighted by the prior,
\be
P(L) = \int df_L\, f_L P(f_L)\,.
\ee
To get the posterior chance, we integrate over our posterior
probability distribution for $f_L$, 
\be
P(L|D) = \int df_L\, f_L P(f_L|D)
 = \frac{\int_0^1 df_L\, f_L P(f_L)/(1+f_L(R-1))}{\int_0^1 df_L P(f_L)/(1+f_L(R-1))}\,.\label{eq.prob_survival}
\ee
Given $R$ and the prior distribution $P(f_L)$,
\eqref{eq.prob_survival} allows us to determine our civilization's
prospects for long-term survival.

Equation \eqref{eq.prob_survival} is exact, but since we are
interested only in $R\gg1$, we can always approximate
\be
P(L|D) \approx
\frac{\int_0^1 df_L\, f_L P(f_L)/(1+f_LR))}{\int_0^1 df_L P(f_L)/(1+f_LR)}
= \frac{\int_0^1 df_L\, f_L P(f_L)/(R^{-1}+f_L)}{\int_0^1 df_L P(f_L)/(R^{-1}+f_L)}\,.
\label{eq.prob_survival2}
\ee
In many cases the contribution from $f_L\lesssim R^{-1}$ to the
numerator is not significant, and we can thus approximate the
numerator by $\int_0^1 df_L P(f_L) = 1$, giving
\be\label{eq.simpler}
P(L|D) \approx
\left[\int_0^1 df_L \frac{P(f_L)}{(R^{-1}+f_L)}\right]^{-1}.
\ee
In some cases there is a cutoff on $f_L$ that keeps it above $R^{-1}$
and in such cases we can write
\be\label{eq.simplest}
P(L|D) \approx
\left[\int_0^1 df_L \frac{P(f_L)}{f_L}\right]^{-1}.
\ee

From \eqref{eq.simplest} we can understand the general effect of the
universal doomsday argument.  The chance that our civilization will
survive to large sizes is small when the integral is large.  This
happens whenever there is a possibility of small $f_L$ where the prior
probability of those $f_L$ is large compared to the $f_L$ themselves.
So, for example, if our prior gives a collective probability of
$10^{-6}$ to a range of $f_L$ near $10^{-9}$, i.e., we think there's
one chance in a million that only a billionth of all civilizations
grow large, then our chance of survival is no more than $10^{-3}$.

The effect of using \eqref{eq.simpler} instead of \eqref{eq.simplest}
is that the above argument does not apply to $f_L$ below $R^{-1}$.
The doomsday argument is able to increase a prior probability
by factor $R$ at most, so scenarios with prior probabilities below 
$R^{-1}$ will never be important.

In the rest of this paper, we will consider several specific priors
and see the conclusions to which they lead.

\section{The uniform prior}
\label{sec:firstprior}

\begin{figure}
\begin{center}
\includegraphics[width=3.5in]{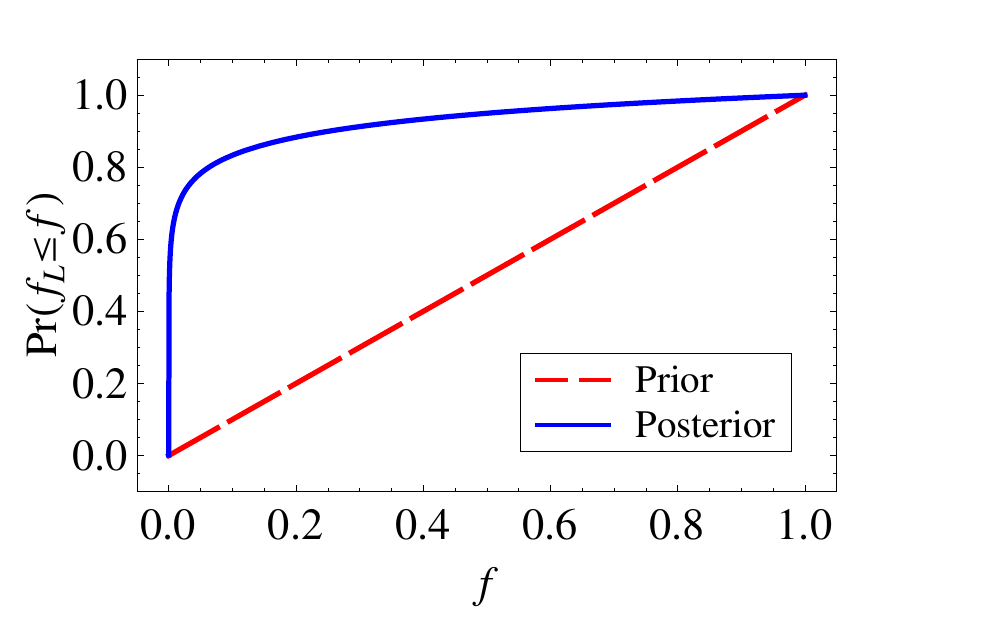}
\end{center}
\caption{Cumulative priors and posteriors with $R=10^6$ and equal prior
  credence given to all $f_L$}
\label{fig.uniform}
\end{figure}

The results for $P(L|D)$ will depend on our prior $P(f_L)$.  Let us
start by taking the simplest prior,
\begin{equation}
P(f_L) = 1\,,
\end{equation}
so that the $f_L$ is equally likely to have any value.  This appears to
be a reasonable choice of prior when we do not have much quantitative
information on the existential threats that we are facing.  The
posterior density is then
\begin{equation}
P(f_L|D) = \frac{R-1}{(\ln R)(1 + f_L(R-1))}\,.
\end{equation}
In Fig.~\ref{fig.uniform}, we plot the cumulative of the prior and
posterior, $P(f_L\le f)$ and $P(f_L\le f|D)$, given $R=10^6$.
Although the probability of low values of $f_L$ increases after
considering $D$, the result is not extreme.  For example, the
probability that $f_L>1/2$ is $5\%$ --- a low but non-negligible amount.

The probability that our civilization is long-lived, after considering $D$, is
\begin{equation}
P(L|D) = \frac{1}{\ln R}-\frac{1}{R-1}\,.
\label{eq.uniformprospects}
\end{equation}
Because $P(L|D)\sim 1/\ln R$, our long-term prospects are not too
bad.  For example, when $R=1$ million, our civilization's chance of
long-term survival is approximately $7\%$.

We can compare these results with the traditional doomsday argument.
Using the uniform prior, the prior chance that our civilization would
be long-lived is $1/2$, and the posterior chance about $1/\ln R$.  If
we took a prior chance of survival $P(L) = 1/2$ in the traditional
doomsday argument, \eqref{eq.traditional} would give our chance of
survival as only $1/(R+1)$.  Thus, at least in this case, taking
account of the existence of multiple civilizations yields a much more
optimistic conclusion.

We can reproduce the traditional doomsday argument even in the
universal setting, merely by asserting that all civilizations have the
same fate, so the benefit of multiple civilizations is eliminated.
This would imply giving no credence to any possibilities except $f_L =
0$ (all civilizations short-lived) and $f_L = 1$ (all civilizations
long-lived).  Thus we could write
\begin{equation}\label{eq.traditionalprior}
P(f_L) = P(L)\delta(f_L) + (1-P(L))\delta(f_L-1)\,.
\end{equation}
Using \eqref{eq.traditionalprior} in \eqref{eq.prob_survival}
reproduces \eqref{eq.traditional}.

\section{N existential threats}
\begin{figure}
\begin{center}
\includegraphics[width=3.5in]{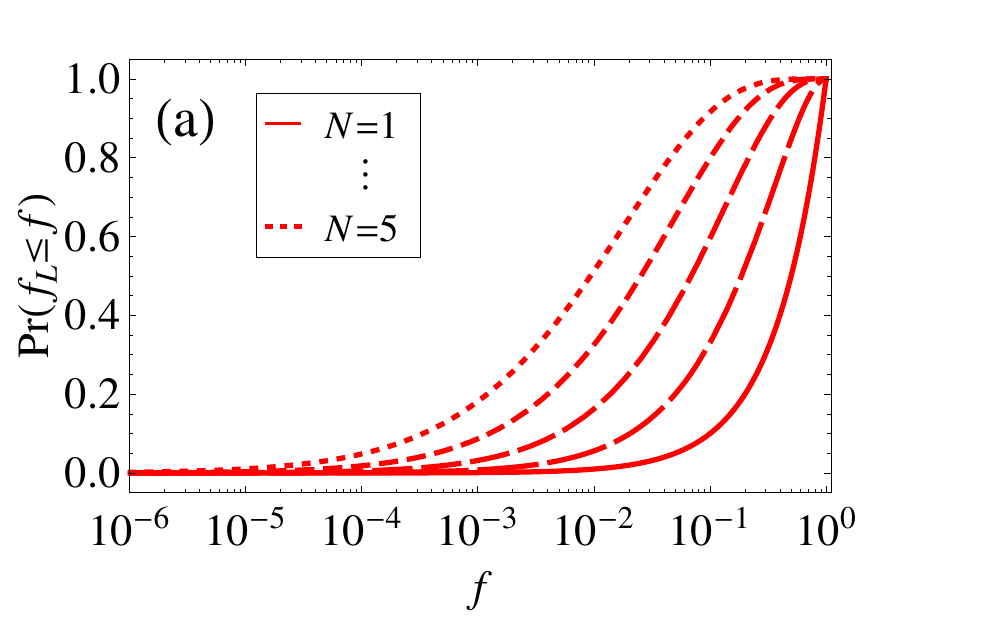}
\includegraphics[width=3.5in]{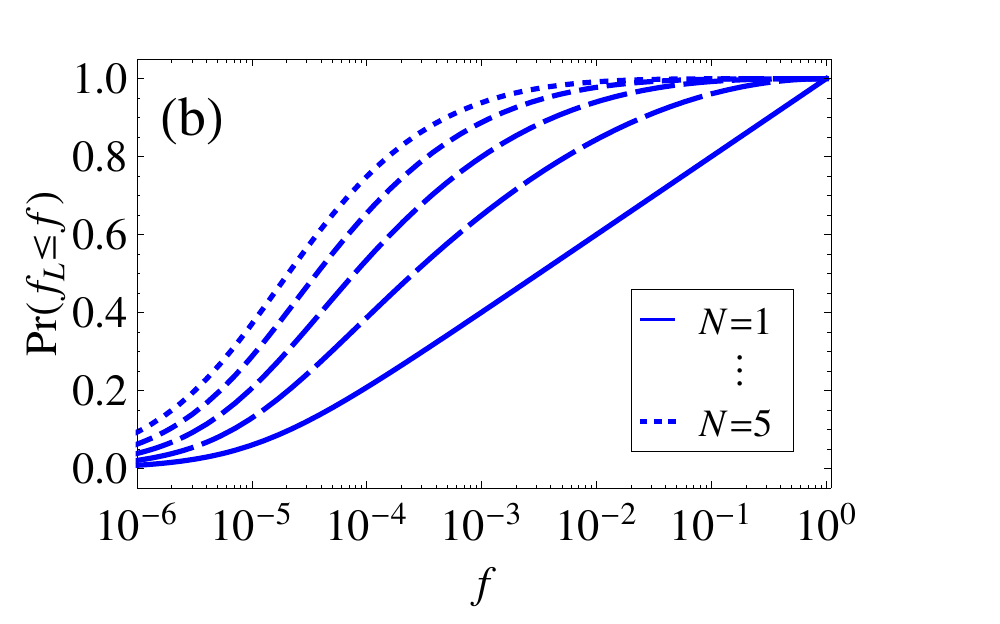}
\end{center}
\caption{The cumulative of the (a) priors and (b) posteriors of $f_L$ given $N$ existential threats and $R=10^6$.}
\label{fig.threats}
\end{figure}

Of course we know that civilizations face more than one existential
threat.  So let us consider the case where there are $N$ statistically
independent threats and take a uniform prior for the fraction of
civilizations that survive each one.  Denote the fraction of
civilizations surviving the $i$-th threat $f_i$.  The fraction of
civilizations that survive all threats is then
\begin{equation}
f_L = f_1 f_2 f_3 \dots f_N\,,
\end{equation}
and our prior for each threat is
\begin{equation}
P(f_i) = 1\,,
\end{equation}
so that
\begin{equation}
P(f_1,f_2,\dots,f_N)=P(f_1)P(f_2)\dots P(f_N)=1\,.
\end{equation}
We can determine the density function for the overall prior, $P(f_L)$,
as follows.  Let $l = |\ln f_L| = -\ln f_L = \sum l_i$ where $l_i =
|\ln l_i|$.  Then
\be
P(l) = P(f_L) \frac{df_L}{dl} = P(f_L) e^{-l}\,.
\ee
Similarly $P(l_i) = e^{-l_i}$, so $l$ is the sum of $N$ independent
and exponentially distributed random variables, and $P(f_L)$ is thus
given by an Erlang (Gamma) distribution,
\be
P(l) = \frac{l^{N-1}e^{-l}}{(N-1)!}\,,
\ee
giving
\begin{equation}
P(f_L) = \frac{|\ln f_L|^{N-1}}{(N-1)!}\,.
\end{equation}
The cumulative of the prior, $P(f_L\le f)$, is shown in
Fig.~\ref{fig.threats}(a)
for $N=1$ to $N=5$.  Note that our civilization's prospects for
long-term survival become rather bleak as $N$ increases.  Even without
considering our datum, $P(L)=1/2^N$.

Now, considering our datum $D$, the posterior density is,
\begin{equation}
P(f_L|D) = \frac{(1-R)}{\Li_N(1-R)}\frac{|\ln f_L|^{N-1}}{(N-1)!(1+f_L(R-1))}\,,
\end{equation}
where $\Li_N$ is the polylogarithm function, given by
\be
\Li_N(z) = \frac{1}{\Gamma(N)}\int_0^\infty
\frac{t^{N-1}}{z^{-1}e^t-1}\,.
\ee
The cumulative of the posterior, $P(f_L\le f|D)$, is shown in
Fig.~\ref{fig.threats}(b) for $N=1$ to $N=5$ and $R=10^6$.

\begin{figure}
\begin{center}
\includegraphics[width=3.4in]{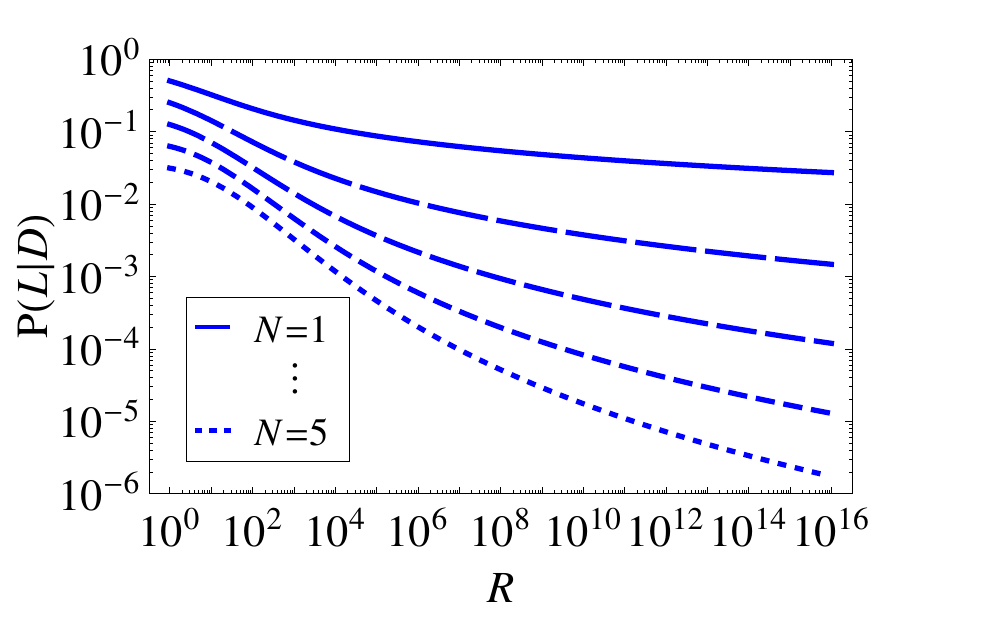}
\end{center}
\caption{The probability that our civilization will be long-lived as a function of $R$ given $N$ existential threats.}
\label{fig.prospects}
\end{figure}

In Fig.~\ref{fig.prospects}, $P(L|D)$ is shown as a function of $R$
for $N=1$ through $N=5$.
After considering $D$, our civilization's prospects for long-term survival are
\begin{equation}
P(L|D)=-\frac{1}{\mathrm{Li}_N(1-R)} - \frac{1}{R-1}\,.
\label{eq.Nprospects}
\end{equation}
When $\ln R\gg N$, we can approximate
\be\label{eq.Nprospects2}
P(L|D)\approx \frac{N!}{(\ln R)^N}\,.
\ee

Notice that when $N=1$, the prior distribution for $f_L$ is uniform
and the results are the same as in the previous section.  As seen in
Fig.~\ref{fig.prospects}, increasing the number of existential
threats, $N$, decreases the probability that our civilization will be
long-lived.  The chance of survival before taking into account $D$ is
$P(L) = 2^{-N}$.  Updating adds the additional factor
\be
\frac{P(L|D)}{P(L)}\approx \frac{2^N N!}{(\ln R)^N}\,,
\ee
which is small whenever $\ln R\gg N$.  However, only a power of $\ln
R$, rather than $R$ itself, appears in the denominator, so the effect is
more benign than in the traditional doomsday argument of
\eqref{eq.traditional}, so long as $N$ is fairly small.

\section{An unknown number of threats: Gaussian distribution}

It would be foolish to imagine that we know of all existential
threats.  For example, before the 1930's no one could have imagined
the threat of nuclear war.  So there is some uncertainty about the
number of threats, $N$.  To make a simple model of this effect, let us
assume that a fixed fraction $q$ of civilizations survive each threat, and
thus $f_N = q^N$ survive them all.  We will take a Gaussian prior for
the distribution of $N$,
\be\label{eq.PNGaussian}
P_N = \frac{1}{Z} e^{-(N/N_0)^2}\,,
\ee
where
\be\label{eq.Zapprox}
Z = \sum_{N=0}^\infty e^{-(N/N_0)^2} \approx \frac{\sqrt{\pi}N_0+1}{2}\,.
\ee
For the purposes of the present section it is a good enough
approximation to drop the $1$ and just use $Z= \sqrt{\pi}N_0/2$.

The prior expectation value for $N$ is
\be
\langle N \rangle\ = \sum N P_N \approx \int_0^\infty N P_N =
\frac{N_0}{\sqrt{\pi}}\,.
\ee
The prior chance that our civilization will grow
 large is
\be\label{eq.GaussianPL}
P(L) = \sum P_N f_N \approx \frac{1}{Z}\sum f_N = \frac{1}{Z(1-q)}\,,
\ee
providing that $N_0|\ln q|\gg 1$, so that the Gaussian does not
decline until $q^N$ is already small.  Thus the prior chance of
survival is quite appreciable.

The posterior chance of survival may be approximated
\be
P(L|D) \approx \left[\sum \frac{P_N}{R^{-1} + f_N}\right]^{-1}
= Z \left[\sum \frac{e^{-N^2/N_0^2}}{R^{-1} + q^N}\right]^{-1}\,.
\ee
As $N$ increases, the numerator in the sum decreases more and more
quickly.  We can approximate that the denominator decreases by a
factor $q$, until $N=\ln R/|\ln q|$, at which point it becomes
constant.  Thus for $N<\ln R/|\ln q|$, the $N$th term in the sum is
larger than the previous term by factor of order
\be
e^{-2N/N_0^2}/q = e^{|\ln q|-2N/N_0^2}\,,
\ee
so the sum is dominated by terms near $N=N_0^2|\ln q|/2$ or near
$N=\ln R/|\ln q|$, whichever is smaller.

In the former case, it is interesting to note that $N$ is proportional
to $N_0^2$, not $N_0$ as one might have thought.  The doomsday argument
partly cancels the Gaussian suppression of the prior probability,
which drops rapidly when $N>N_0$.

We can approximate the sum by an integral,
\be
\int_0^\infty dN\, \exp\left( -\frac{(N-N_0^2|\ln q|/2)^2}{N_0^2}
+\frac{N_0^2|\ln q|^2}{4}\right)
\approx \sqrt{\pi}N_0\exp\left(\frac{N_0^2|\ln q|^2}{4}\right)\,,
\ee
so
\be\label{eq.variable}
P(L|D) \approx \frac{1}{2} \exp\left(-\frac{N_0^2|\ln q|^2}{4}\right)
\approx \frac{1}{2}e^{-0.12 N_0^2}\,,
\ee
where the last step is for $q=1/2$.  If $N_0$ is small, this is not
too pessimistic a conclusion.  For example with $N_0=5$ threats we
find $P(L|D) \approx 0.02$, but for $N_0=10$ we find $P(L|D) \approx
2\times 10^{-6}$.

To be in this regime, we require that  $N_0^2|\ln q|/2 < \ln R/|\ln
q|$, i.e.,
\be
\ln R > \frac{N_0^2|\ln q|^2}{2} \approx 0.24 N_0^2\,,
\ee
for $q=1/2$.  For $N_0 =10$ this requires $R>2\times 10^{10}$, which
is not unreasonable.

\begin{figure}
\begin{center}
\includegraphics[width=3.4in]{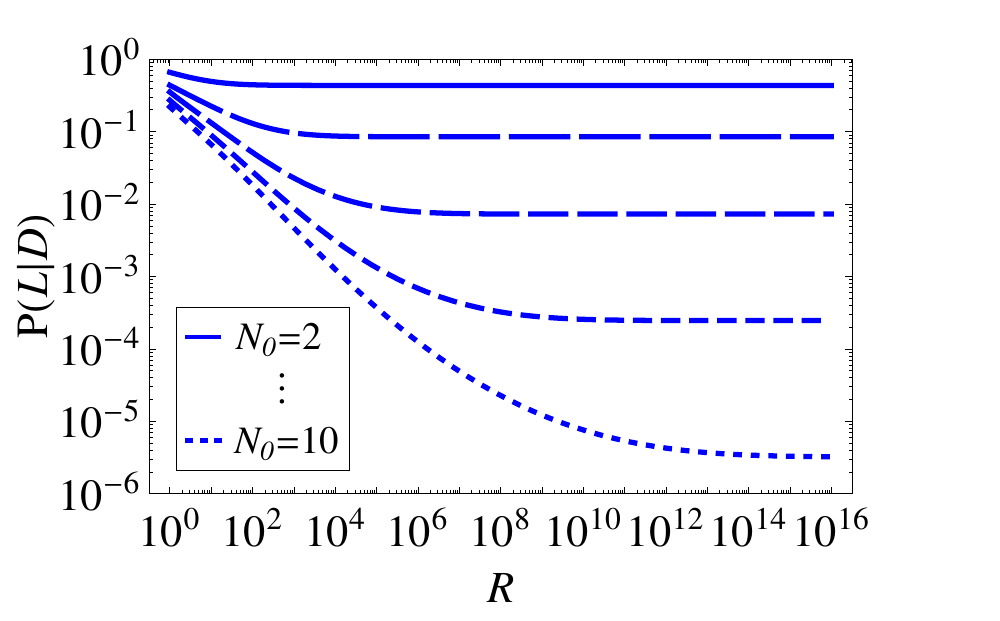}
\end{center}
\caption{The probability that our civilization will be long-lived as a
  function of $R$ with the Gaussian prior of \eqref{eq.PNGaussian} for
  the number of threats, with the parameter $N_0=2,4,6,8,10$.}
\label{fig.Gaussianprospects}
\end{figure}

Fig.~\ref{fig.Gaussianprospects}
 shows our prospects for long-term
survival as a function of $R$ for various values of $N_0$.  Because
$\langle N \rangle \approx N_0/\sqrt{\pi} \approx N_0/2$, a given
$N_0$ corresponds roughly to choosing $N=N_0/2$ in the previous section.

Comparing Fig.~\ref{fig.Gaussianprospects} for each $N_0$ to
Fig.~\ref{fig.prospects} for $N=2 N_0$, we see that, as $R$ becomes
larger, Fig.~\ref{fig.prospects} shows lower survival
chances than Fig.~\ref{fig.Gaussianprospects}.  In the present case the
doomsday argument amplifies the probability of small $f$ that arise
from a large number of threats, but because we took fixed $q$ we don't
consider threats that have very low survival probability.  In the
previous section, we fixed $N$, but the uniform prior gave some
chance to arbitrarily small $f$.

\section{An unknown number of threats: exponential distribution}
\label{sec:lastprior}

The reason that the conclusion of the previous section was not too
pessimistic is that the Gaussian prior for $P_N$ very strongly
suppressed numbers of threats $N\gg N_0$.  If we take a distribution
that falls only exponentially, the result may be quite different.  So
consider now
\be\label{eq.PNexponential}
P_N = (1-s) s^N\,,
\ee
for some $s<1$.  The prior expectation value for $N$ is
\be
\langle N \rangle\ = \sum N P_N = \frac{s}{1-s}\,,
\ee
and the prior chance of survival is
\be
P(L) = \sum P_N f_N = \frac{1-s}{1-sq}\,.
\ee
The posterior chance of survival is given by
\be\label{eq.variableexp}
P(L|D) \approx \left[(1-s)\sum \frac{s^N}{R^{-1} + q^N}\right]^{-1}\,.
\ee
If $s<q$, our prior credence in $N$ threats decreases faster than the
chance of surviving $N$ threats, and the terms in the sum in
\eqref{eq.variableexp} decrease.  Except when $q$ is very close to
$s$, we can ignore $R$ and get an optimistic conclusion
\be
P(L|D) \approx \frac{1-s/q}{1-s}\,.
\ee
But if $s>q$, the situation is different.  In this case, our credence
in a large number of threats is higher than the survival chance, and
then the universal doomsday argument acts to increase our prior
for facing many threats and so decrease our expectation of survival.
The terms in the sum in \eqref{eq.variableexp} increase as $(s/q)^N$
until we reach some $N$ where the existence of $R$ begins to
matter.  This happens when $q^N\sim R^{-1}$, i.e., when
\be\label{eq.Nprime}
N\approx N' = \frac{\ln R}{|\ln q|}\,.
\ee
For $N>N'$ we can ignore $q$ to get $R s^N$ in the sum of
\eqref{eq.variableexp}.  Thus we can split our sum into two parts,
which give
\be
\frac{(s/q)^{N'}-1}{s/q-1} + \frac{R s^{N'}}{1-s}
= \left(\frac{1}{s/q-1}+\frac{1}{1-s}\right)R^{1-|\ln s|/|\ln q|}\,,
\ee
and so
\be\label{eq.expPLD}
P(L|D) \approx \frac{s-q}{s-sq} R^{|\ln s|/|\ln q|-1}\,.
\ee
As long as $s$ is significantly larger than $q$, this is very small.
As an example, we can choose $q=1/2$ and $s=3/4$.  This gives the
prior probability $P(L) = 0.8$ here, while \eqref{eq.GaussianPL} and
\eqref{eq.Zapprox} give $P(L) = 0.81$.  While these are nearly the
same, \eqref{eq.expPLD} with $R=10^9$ gives
\be
P(L|D) \approx 4 \times 10^{-6}\,,
\ee
must lower than $P(L|D) \approx 0.02$ from \eqref{eq.variable}.

\begin{figure}
\begin{center}
\includegraphics[width=3.4in]{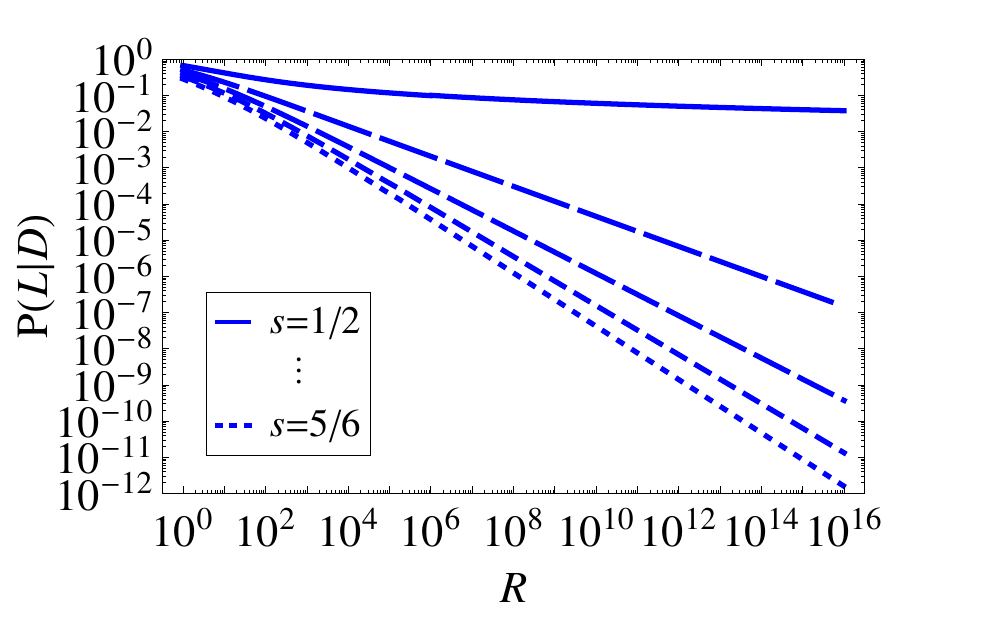}
\end{center}
\caption{The probability that our civilization will be long-lived as a
  function of $R$ with the exponential prior of \eqref{eq.PNexponential} for
  the number of threats, with the parameter $s=1/2,2/3,3/4,4/5,5/6$
  corresponding to $\langle N\rangle$ being $1,2,3,4,5$.}
\label{fig.exponentialprospects}
\end{figure}

Fig.~\ref{fig.exponentialprospects} shows our prospects for long-term
survival as a function of $R$ for various values of $s$.

\section{Summary and discussion}
\label{sec:summary}

Bayes' Theorem tells us how the probabilities we assign to various
hypotheses should be updated when we learn new information.  The
Doomsday Argument is concerned with the impact of the important piece
of information that we called $D$ --- that we are among the first $N_S$
humans to be born.  Earlier investigations
\cite{Carter:unpublished,Leslie:1989,Gott:doomsday,Nielsen:doomsday,Knobe:2003js,Gerig:2012qg}
suggested that the resulting probability for our civilization to be
long-lived is suppressed by a huge factor $R=N_L/N_S\gg 1$, where
$N_L$ is the size a civilization may reach if it does not succumb to
early existential threats.  Here, we attempted a more careful analysis
by considering a number of possible choices of prior probabilities.
We found that, with a seemingly reasonable choice of the prior, our
chances of long-term survival are suppressed by a power of $\ln R$,
rather than by $R$, with the power determined by the number of threats
$N$.  If $N$ is not too large, the probability of long-term survival
is about a few percent.

This conclusion has been reached by assuming a flat prior
distribution, $P(f_i)=1$, for the fraction of civilizations $f_i$
surviving statistically independent threats (labeled by $i$).  This
appears to be a reasonable assumption, reflecting our present state of
ignorance on the subject.  We also considered a prior where the
survival probability for each threat is a fixed number $q$, while the
number of threats $N$ is a stochastic variable with a Gaussian
distribution of width $N_0$.  Once again, we find, assuming that $q$
is not too small and $N_0$ is not too large, that the Bayesian
suppression factor is much smaller than suggested by the naive
Doomsday Argument.
  
In our analysis, we adopted a model where civilizations can have only
two possible sizes, $N_S$ and $N_L$.  This model is of course not
realistic in detail, but it may well capture the bimodal character of
the realistic size distribution.  Civilizations that actively engage
in colonizing other planetary systems and reach a certain critical
size are likely to grow extremely large, while civilizations confined
to their home planet must be rather limited in size.

Even though we found a greater survival probability than in
Refs.~\cite{Carter:unpublished,Leslie:1989,Gott:doomsday,Nielsen:doomsday,Knobe:2003js,Gerig:2012qg},
our conclusions can hardly be called optimistic.  With the priors that
we considered, the fraction of civilizations that last long enough to
become large is not likely to exceed a few percent.  If there is a
message here for our own civilization, it is that it would be wise to
devote considerable resources (i) for developing methods of diverting
known existential threats and (ii) for space exploration and
colonization.  Civilizations that adopt this policy are more likely to
be among the lucky few that beat the odds.  Somewhat encouragingly,
our results indicate that the odds are not as overwhelmingly low as
suggested by earlier work.

\appendix
\section{Arbitrary possibilities for civilization size}
\def\vf{\mathbf{f}}
\def\A{\langle A\rangle}

In this appendix we give a formalism for considering all possible
civilization sizes.  Let a scenario be a set of numbers $f_n$, $n =
0\ldots\infty$ giving the fraction of civilizations that have each
size $n$.  Since every civilization has some size, we must have
$\sum_n f_n = 1$.  We will write $\vf$ for the entire vector of
numbers $f_n$.  The average number of observers per civilization in
scenario $\vf$ is
\be
n(\vf) = \sum_n n f_n\,,
\ee
which we will assume is finite.

Now let $P(\vf)$ denote the prior probability that we assign to each
possible scenario $\vf$.  The probabilities must be normalized,
\be
\int d\vf\, P(\vf) = 1\,,
\ee
where
\be
\int d\vf\,\quad \hbox{denotes}\quad\int_0^1 df_0\int_0^1 df_1\int_0^1
df_2 \ldots.
\ee
We suppose that $P(\vf)$ already contains a term such as
$\delta(1-\sum_n f_n)$ that excludes unnormalized $\vf$.

Let $n_0$ denote the size of our civilization.  We will not be
concerned here with issues related to civilizations with fewer than
$n_0$ observers, so let us suppose that $P(\vf)$ is supported only
when $f_n=0$ for $n<n_0$.  Now we consider the datum $D$, that we are
in the first $n_0$ individuals in our civilization.  The chance for a
randomly chosen observer to observe $D$ is
\be
P(D|\vf) = \frac{n_0}{n(\vf)}\,.
\ee
Now let $A$ be some property of $\vf$, such as the average size of a
civilization or the chance that the civilization has more than a
certain number of members.  The average value of $A$ not taking into
account $D$ is
\be
\A = \int d\vf\, A(\vf) P(\vf)\,.
\ee

Now we take $D$ into account using Bayes' Rule.  We find
\be\label{eq.appendix_posterior}
P(\vf|D) = \frac{P(D|\vf)P(\vf)}{\int d\vf'\, P(D|\vf')P(\vf')}
= \frac{P(\vf)/n(\vf)}{\int d\vf'\,P(\vf')/n(\vf')}\,,
\ee
which is the arbitrary-size generalization of \eqref{eq.posterior}.
The average value of $A$ taking into account $D$ is
\be
\A|D = \int d\vf\,A(\vf)P(\vf|D) = \frac{\int d\vf\,A(\vf)P(\vf)/n(\vf)}
{\int d\vf\, P(\vf)/n(\vf)}\,.
\ee

A particularly simple case is when $A(\vf) = n(\vf)$.  The expected
value of the total size of our civilization taking into account
$D$ is
\be
\langle n\rangle|D = \frac{1}{\int d\vf\, P(\vf)/n(\vf)}\,.
\ee

Alternatively, let $f_L(\vf) = \sum_{n=N_L}^\infty f_n$ be the
fraction of civilizations in scenario $\vf$ that grow larger than some
threshold $N_L$.  The posterior chance that our civilization will
reach this threshold is then
\be\label{eq.fL}
P(L|D) = \langle f_L \rangle|D = \frac{\int d\vf\,f_L(\vf)P(\vf)/n(\vf)}
{\int d\vf\, P(\vf)/n(\vf)}\,.
\ee
This is just the fraction of large civilizations in the different
scenarios, weighted by prior probability of the scenario and the
inverse of the average civilization size according to that scenario.

Unfortunately, the set of possible priors is so large here that it is
difficult to make any progress.  It seems likely to us that
civilizations will either remain confined to a single planet and
eventually be wiped out by some disaster, or spread through the galaxy
and grow to large size.  We can approximate this by considering
$P(\vf)$ supported only at two sizes $N_S$ and $N_L$,
\be
P(\vf)=P(f_L)\, \delta(f_S+f_L-1)\prod_{n\neq N_L, N_S} \delta(f_n)\,,
\ee
where we have written $f_L$ for $f_{N_L}$ and $f_S$ for $f_{N_S}$.
Then we recover the results in the main text. Integrals $P(\vf) d\vf$
become $P(f_L) df_L$, $n(\vf)$ is just $(1-f_L)N_S+ N_Lf_L$,
$f_L(\vf)$ is $f_L$, \eqref{eq.appendix_posterior} becomes
\eqref{eq.posterior}, and \eqref{eq.fL} becomes
\eqref{eq.prob_survival}.

\bibliographystyle{JHEP}

\providecommand{\href}[2]{#2}\begingroup\raggedright\endgroup

\end{document}